\documentclass[usegraphicx,usenatbib,useAMS,twocolumn]{mn2e}
\usepackage{epsfig}
\usepackage{graphicx}
\usepackage{amsmath,bm}
\usepackage{color}
\usepackage{amssymb}
\usepackage{textcase}

\newcommand{\combo}{{$f^{0.43}\sigma_8|_{z=0.57}$ }}

\newcommand{\combonospace}{{$f^{0.43}\sigma_8|_{z=0.57}$}}

\title[The power spectrum and  bispectrum of SDSS DR11 BOSS galaxies II: cosmological interpretation]{The power spectrum and  bispectrum of SDSS DR11 BOSS galaxies II: cosmological interpretation}
\author[H. Gil-Mar\'in et al.]{H\'ector Gil-Mar\'in$^{1}$\thanks{hector.gil@port.ac.uk},  Licia Verde$^{2,3,4}$, Jorge Nore\~na$^{2,5}$, Antonio J. Cuesta$^2$ , \and  Lado Samushia$^{1}$, 
Will J. Percival$^{1}$, Christian Wagner$^{6}$, Marc Manera$^7$, \and Donald P. Schneider$^{8,9}$\\
 $^{1}$ Institute of Cosmology \& Gravitation, University of Portsmouth, Dennis Sciama Building, Portsmouth PO1 3FX, UK\\
 $^{2}$ Institut de Ci\`encies del Cosmos, Universitat de Barcelona, IEEC-UB, Mart\'i i Franqu\`es 1, 08028, Barcelona, Spain\\
    $^{3}$ ICREA (Instituci\'o Catalana de Recerca i  Estudis Avan\c{c}ats), Passeig Llu\'is Companys, 23 08010 Barcelona - Spain\\
 $^{4}$ Institute of Theoretical Astrophysics, University of Oslo, Norway\\
  $^{5}$ Department of Theoretical Physics and Center for Astroparticle Physics (CAP), 24 quai E. Ansermet, CH-1211 Geneva 4, CH \\
 $^{6}$ Max-Planck-Institut f\"ur Astrophysik, Karl-Schwarzschild Str. 1, 85741 Garching, Germany\\
  $^{7}$ University College London, Gower Street, London WC1E 6BT, UK\\
    $^8$ Department of Astronomy and Astrophysics, The Pennsylvania State University, University Park, PA 16802, USA\\
$^9$ Institute for Gravitation and the Cosmos, The Pennsylvania State University, University Park, PA 16802, USA\\
  }
  
\def\gs{\mathrel{\raise1.16pt\hbox{$>$}\kern-7.0pt
\lower3.06pt\hbox{{$\scriptstyle \sim$}}}}         
\def\ls{\mathrel{\raise1.16pt\hbox{$<$}\kern-7.0pt 
\lower3.06pt\hbox{{$\scriptstyle \sim$}}}}         

\begin{document}
\maketitle

\begin{abstract}
We examine the cosmological implications of the measurements of  the linear growth rate of cosmological structure obtained in a companion paper from the power spectrum and bispectrum monopoles of the Sloan Digital Sky Survey III Baryon Oscillation  Spectroscopic Survey  Data, Release 11, CMASS  galaxies. This measurement was of $f^{0.43}\sigma_8$,  where $\sigma_8$ is the amplitude of dark matter density fluctuations,  and $f$ is the linear growth rate, at the effective redshift of the survey, $z_{\rm eff}=0.57$. In conjunction with Cosmic Microwave Background (CMB) data, interesting constraints can be placed on models with non-standard neutrino properties and models where gravity deviates from  general  relativity on cosmological scales.  In particular,  the sum of the masses of the three species of the neutrinos is constrained to  $m_\nu<0.49\,{\rm eV}$ (at 95\% confidence level) when the $f^{0.43}\sigma_8$ measurement is combined with  state-of-the-art CMB measurements. Allowing the effective number of neutrinos to vary as a free parameter does not significantly change these results. When we combine the measurement of $f^{0.43}\sigma_8$ with the complementary measurement of $f\sigma_8$ from the monopole and quadrupole of the two-point correlation function we are able to obtain an independent measurements of $f$ and $\sigma_8$. We obtain $f=0.63\pm0.16$ and $\sigma_8=0.710\pm0.086$ (68\% confidence level). This is the first time when these parameters have been able to be measured independently using the redshift-space power spectrum and bispectrum measurements from galaxy clustering data only.

\end{abstract}


\section{introduction}
\label{sec:1}
Direct and model-independent constraints on the growth of cosmological structure  are particularly important in cosmology. Measurements of the expansion history of the Universe (via standard candles or standard rulers) have clearly indicated an accelerated expansion since redshift $z\sim 0.3$, but are insufficient to  identify the physics causing it; information on the growth of structure is key (for a review of the state of the field and general introduction to it see e.g., \cite{DETF, Euclid,Snowmass}  and references therein).  In particular,  while the cosmological constant remains  at the core of the standard cosmological model, and  is the most popular  explanation for the accelerated expansion, it raises several problems from its smallness (the cosmological constant problem) to its fine tuning (the coincidence problem). 
This situation has led many scientists to investigate  alternatives to vacuum energy (dark energy) or to challenge one of the basic tenets of  cosmology, namely General Relativity (GR). 
 After all, precision tests of GR have been performed on solar system scales, but more than 10 orders of magnitude extrapolation is required to apply it on cosmological scales.  Should GR be modified on cosmological scales, it could still mimic the $\Lambda$CDM expansion history but the growth of structure would be affected.

Most of the  information  we can gather  about clustering and large scale structure, which on large scales would trace the linear growth of perturbations,  come from  galaxies.   It is well known that different kinds  of galaxies show different clustering properties,  and thus not all objects can be  faithful tracers of the underlying  mass distribution; this feature is called galaxy bias. 
There are two notable  observational techniques that avoid  this limitation: gravitational lensing and  redshift-space distortions. Gravitational lensing is an extremely promising approach which, however, at present reaches  limited signal-to-noise ratio in the linear  or mildly non-linear regime.  
The study of redshift-space distortions observed in galaxy redshift surveys  uses galaxies as test particles  in the velocity field and thus this technique is relatively insensitive to galaxy bias \footnote{This technique would be sensitive, of course, to a velocity bias if tracers did not to statistically represent the distribution of velocities of dark matter. Such a velocity bias is not expected on large-scales.}.

A third  approach is to use higher-order correlations to disentangle the effects of gravity from those of galaxy bias (e.g., \citealt{Fry:1994}). This is the approach recently  pursued   in  \cite{paper1} where, by performing a joint analysis of the monopole power spectrum and bispectrum of the CMASS sample of the Baryon Oscillation Spectroscopic Survey Data Release 11 (BOSS DR11 SDSSIII) survey \citep{Gunn98,Gunn06,Eisenstein11,Bolton12,Dawson13,Smee13}, constraints on both galaxy bias and growth  of structures at the effective redshift of the survey $z=0.57$ were obtained.

Here we investigate the cosmological implications of the measurement of the quantity $f^{0.43}\sigma_8$ at $z=0.57$ (hereafter \combo) obtained in \cite{paper1} (hereafter Paper~I). The parameter $f$ quantifies the linear growth rate of structures: $f=d\ln\delta/d\ln a$, where $a$ is the scale factor and $\delta$ the (linear) matter overdensity fluctuation; $\sigma_8$ is the {\it rms} of the (linear) matter over density field extrapolated at $z=0$ and smoothed on scales of $8\,h^{-1} {\rm Mpc}$. 
Paper~I reported \combo$=0.582\pm 0.084$ ($0.584\pm 0.051$); where the first result correspond to the conservative analysis while the second to the more optimistic analysis (see \cite{paper1} for details). Hereafter we report in parenthesis the results  corresponding to the optimistic analysis. The difference in the central values is negligible at all effects thus we will only report the error-bars corresponding to the optimistic analysis.  This 14\% (9\%) error  on the quantity \combo is comparable to that obtained in the quantity $f\sigma_8$ from the study of  redshift-space distortions of the power spectrum of the same survey (e.g., \citealt{Florian,Albert,Samushiaetal:14} and \citealt{Ariel}), which have error-bars of typically $10\%$.  While not being statistically independent (the survey is the same),  the method is complementary and the measurement relies on a different physical effect,  harvesting the power of higher-order correlations rather than the anisotropy of clustering induced by the redshift-space distortions. 
This paper is organised as follows: in \S~\ref{sec:2} we present the data sets we use and the methodology and the consistency  of the measurement with the $\Lambda$CDM model with GR. \S~\ref{sec:3} presents tests constraining several extensions of this model which involve  changes in the  composition (i.e., neutrino properties) or background (i.e., dark energy models and geometry)  or deviations from GR. We explore the potential of combining  the bispectrum monopole  with anisotropic  clustering of the two point function in \S~\ref{sec:PoP} and conclude in \S~\ref{sec:5}.

\section{methods and data sets}
\label{sec:2}

In Paper~I we have analysed the anisotropic clustering of the Baryon Oscillation Spectroscopic Survey (BOSS) CMASS Data Release 11 sample, composed of 690827 galaxies in the redshift range of $0.43<z<0.70$. This surveys covers an angular area of 8498 deg$^2$, which corresponds to an effective volume of $\sim6\,{\rm Gpc}^3$. We have measured the corresponding  redshift-space galaxy power spectrum and bispectrum monopoles, providing  a measurement of the linear growth rate, $f$, in combination with the amplitude of matter density fluctuations, $\sigma_8$, $f^{0.43}\sigma_8=0.582\pm0.084$ ($0.584\pm 0.051$)  at the effective redshift of the survey, $z_{\rm eff}=0.57$.  The optimistic estimate is obtained by pushing slightly more into the mildly non-linear regime and thus  including significantly  more modes. For this particular combination of $f$ and $\sigma_8$,   close to the maximum likelihood solution the likelihood surface  is much closer to that of Gaussian distribution than in the  individual parameters. In addition,  this measurement  is insensitive to the  fiducial cosmology assumed in the analysis. Our measurements are supported by a series of tests performed on dark matter N-body simulations, halo catalogues (obtained both from \textsc{PThalos} and N-body simulations) and mock galaxy catalgoues (see \S~5 in Paper I for an extensive list of tests to check for systematic errors and to asses the performance of the different approximations adopted). These tests are used to identify the regime of validity of the adopted modelling, which occurs when all the $k$-modes of the bispectrum triangles are larger than $0.03h{\rm Mpc}^{-1}$ and less than $0.17h{\rm Mpc}^{-1}$ (with the conservative analysis) or less than $0.20h{\rm Mpc}^{-1}$ being more optimistic. We have also accounted for  real world effects such as the survey window and systematic weighting of objects. We have opted to add in quadrature the statistical error and half of the systematic shift in order to account for the uncertainty in the systematics correction. Because the bispectrum calculation is computationally intensive, we have only considered a subset of all possible bispectrum shapes: $k_2/k_1=1$ and $k_2/k_1=2$. The statistical error on \combo has been obtained from the scatter among 600 mock catalogs.
 Our cosmologically interesting parameters are: the linear matter clustering amplitude $\sigma_8$ and the growth rate of fluctuations $f=d\ln \delta / d \ln a$. In Paper I we showed that even jointly, the power spectrum and bispectrum monopole cannot measure these two parameter separately, but do constrains on the \combonospace. In this variable, the distribution of the best-fit parameters for the galaxy mock catalogues is much closer to a Gaussian distribution than in the separate $\sigma_8$ and $f$ parameters.

We combine the  \combo measurement with the constraints from Cosmic Microwave Background (CMB) observations  acquired  by the {\it Planck} satellite \citep{Planckmission,Planckoverview,Planck_cosmology}. In many cases  we use  the outputs of their Monte Carlo Markov Chains   for importance sampling;  when specified we run new Markov chains.  We use either the {\it Planck +WP} data --{\it Planck} primary temperature data with the Wilkinson Microwave Anisotropy Probe WMAP \citep{WMAP} polarisation data \citep{Bennett:2012fp,Hinshaw:2012aka} at low multipoles---  or  {\it Planck +WP+highL} -- the above data with the addition of high multipoles temperature  observations  from the Atacama Cosmology Telescope  \citep{ACT} and the South Pole telescope \citep{SPT}.   The {\it Planck} maps have also been analysed to extract the weak gravitational lensing signal arising from intervening large scale structure \citep{Plancklensing}. This task is done through the four point function of the temperature maps; when including this information    we refer to it as {\it lensing}.
In some cases the CMB constraints are improved by the addition of information on the  expansion history via Baryon Acoustic Oscillation (BAO) measurements \citep{Beutler:2011hx,Blake:2011en,Padmanabhan:2012hf,Anderson:2012sa,Percival:2010}.

We  investigate whether the \combo measurement is consistent with the $\Lambda$CDM+GR model prediction when the model parameters  are constrained by CMB observations.
\begin{figure}
\center
\includegraphics[trim= 30 0 0 0,clip,scale=0.5]{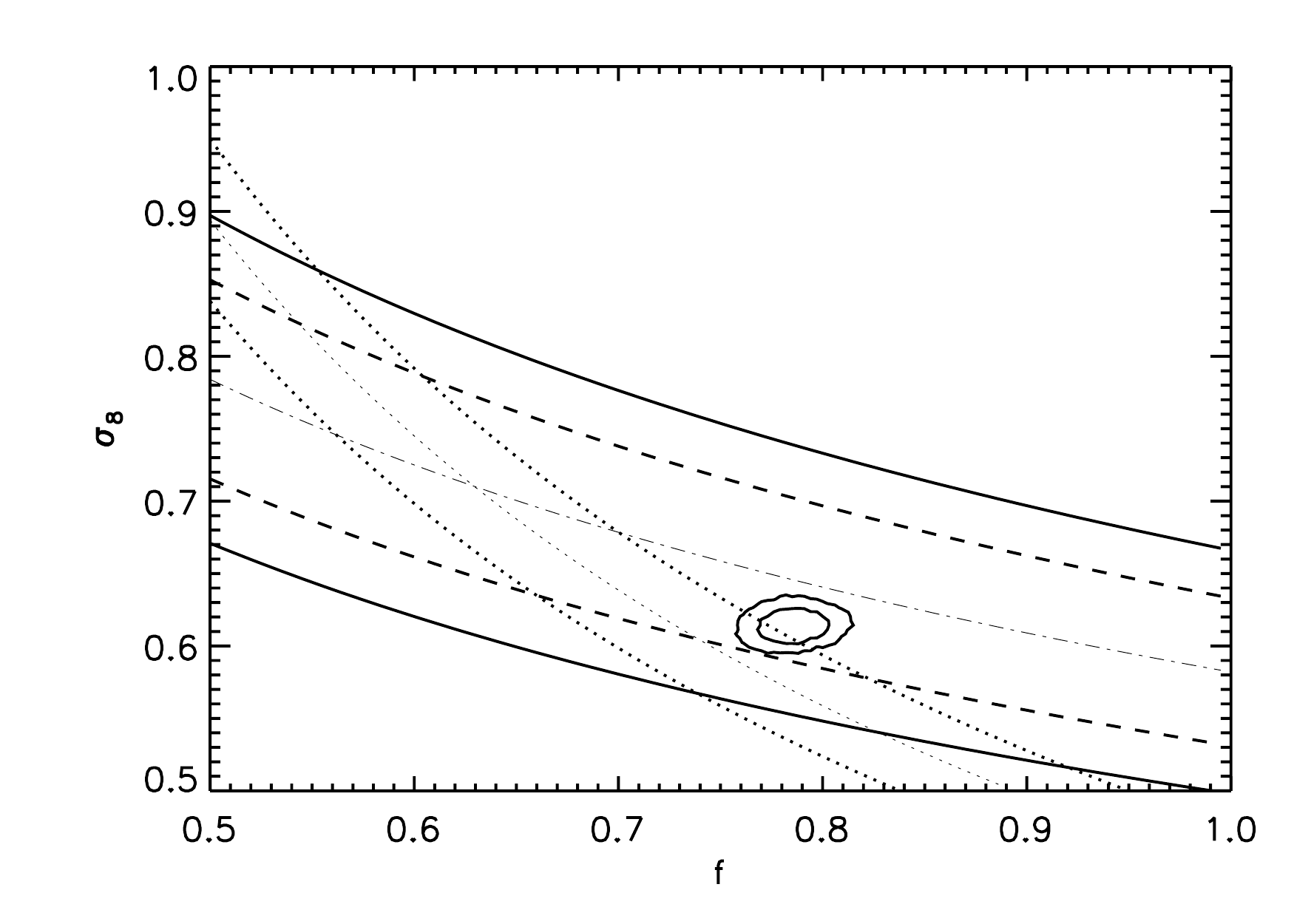}
\caption{Constraints in the $f$-$\sigma_8$ plane where both quantities are at the effective redshift of the BOSS CMASS DR11 galaxies $z=0.57$. The ellipses represent the {\it Planck} CMB inferred constraints  (68 and 95\% confidence) when assuming GR and a $\Lambda$CDM model. The dot-dashed line represents the best-fitting value for the measurements of Paper~I, which uses the monopole power spectrum and bispectrum of the BOSS CMASS DR11 galaxies. The solid and dashed lines represent the 68\% confidence region corresponding to the conservative and optimistic analysis respectively.
 The dotted lines  are the 68\% constraints obtained from the monopole and quadrupole of the two-point function from the same galaxy catalog \citep{Samushiaetal:14}.}
\label{fig1}
\end{figure}
Assuming GR, the expansion history uniquely predicts the linear growth rate.  In a $\Lambda$CDM+GR model, observations of the CMB impose tight constraints on the expansion history and therefore on the growth of structure.

 Fig.~\ref{fig1}  presents the constraints on the $f$--$\sigma_8$-plane,  at $z=0.57$, obtained from  the {\it Planck+WP}  CMB observations  extrapolated assuming GR and a flat-$\Lambda$CDM model, from the direct measurement of Paper~I and,  for completeness, from  the  BOSS CMASS galaxies anisotropic clustering \citep{Samushiaetal:14}: $f\sigma_8= 0.441\pm 0.044$, at $z=0.57$. The  galaxy clustering  measurements are  individually in agreement  with the standard $\Lambda$CDM cosmological paradigm  within $\sim1\sigma$. 

As  the two  measurements constrain different combinations of $f$ and $\sigma_8$,  they can be  measured separately from a combined analysis. We explore this prospect in \S~\ref{sec:PoP}.

Although there is no evidence for significant tensions between the CMB and the lower redshift measurements (in particular the \combo measurement of \cite{paper1}), we consider  standard $\Lambda$CDM model extensions,  where one or more extra cosmological parameters are allowed to vary. We then consider direct constraints on modifications of GR.

In \S~\ref{sec:PoP} we also consider the measurement of the combination $f\sigma_8$ from \citet{Samushiaetal:14}. This reference uses the same data set as Paper~I but exploits the fact that redshift-space distortions  cause an isotropic two-point correlation function  become anisotropic.  
The magnitude of the large-scale velocity field traced by galaxies depends on the nature of gravitational interactions, thus the angular dependence of the two-point function can be used to measure the combination $f\sigma_8$.
For more details see \cite{Reidetal2012, Samushiaetal:14} and references therein.

\section{Results}
\label{sec:3}
In this section we present the constrains that the \combo measurement provide on different extensions of $\Lambda$CDM+GR such as, i) neutrino mass properties, ii) changes in the dark energy equation of state, iii) deviations from GR.

\subsection{Neutrino mass constraints}
\label{sec:3.1}
Among  possible  $\Lambda$CDM  model extensions,  which still assume GR,  we expect  the \combo measurement to provide significant improvement over CMB data alone for the cases of  
 where significant evolution in the growth rate to low redshift is expected. This is the case for massive neutrinos and  for models where dark energy deviates from a cosmological constant and where more than one  evolution-affecting parameter is added to the ``base" $\Lambda$CDM model. In other model extensions, the addition of the growth constraints only reduces the CMB error-bars by few percent. In these cases therefore, this combination offers  a test of consistency   rather than a technique of reducing parameter degeneracies and improving cosmological constraints.
\begin{figure}
\center
\includegraphics[trim= 20 0 0 0,clip, scale=0.5]{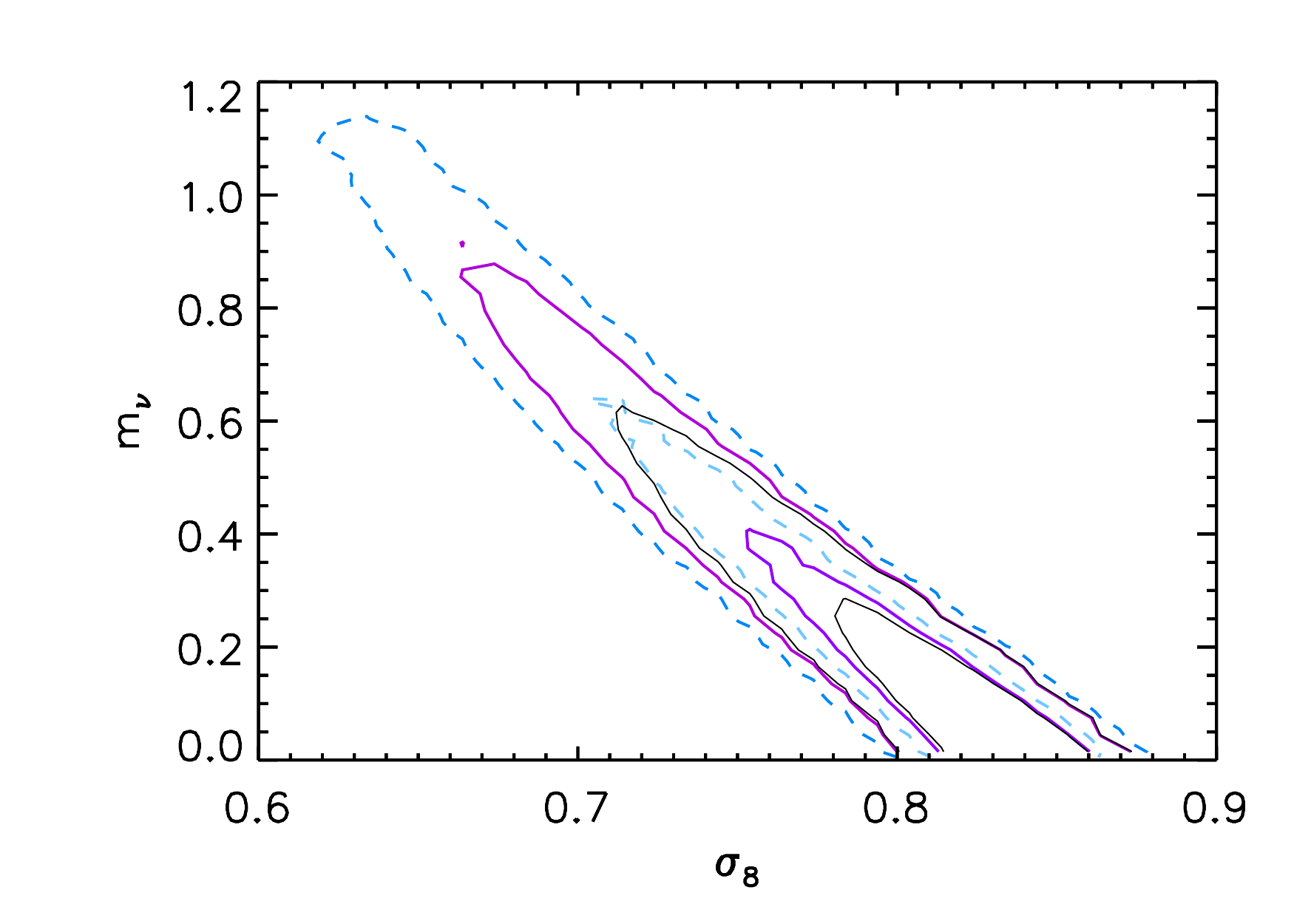}
\includegraphics[trim= 20 0 0 0,clip,scale=0.5]{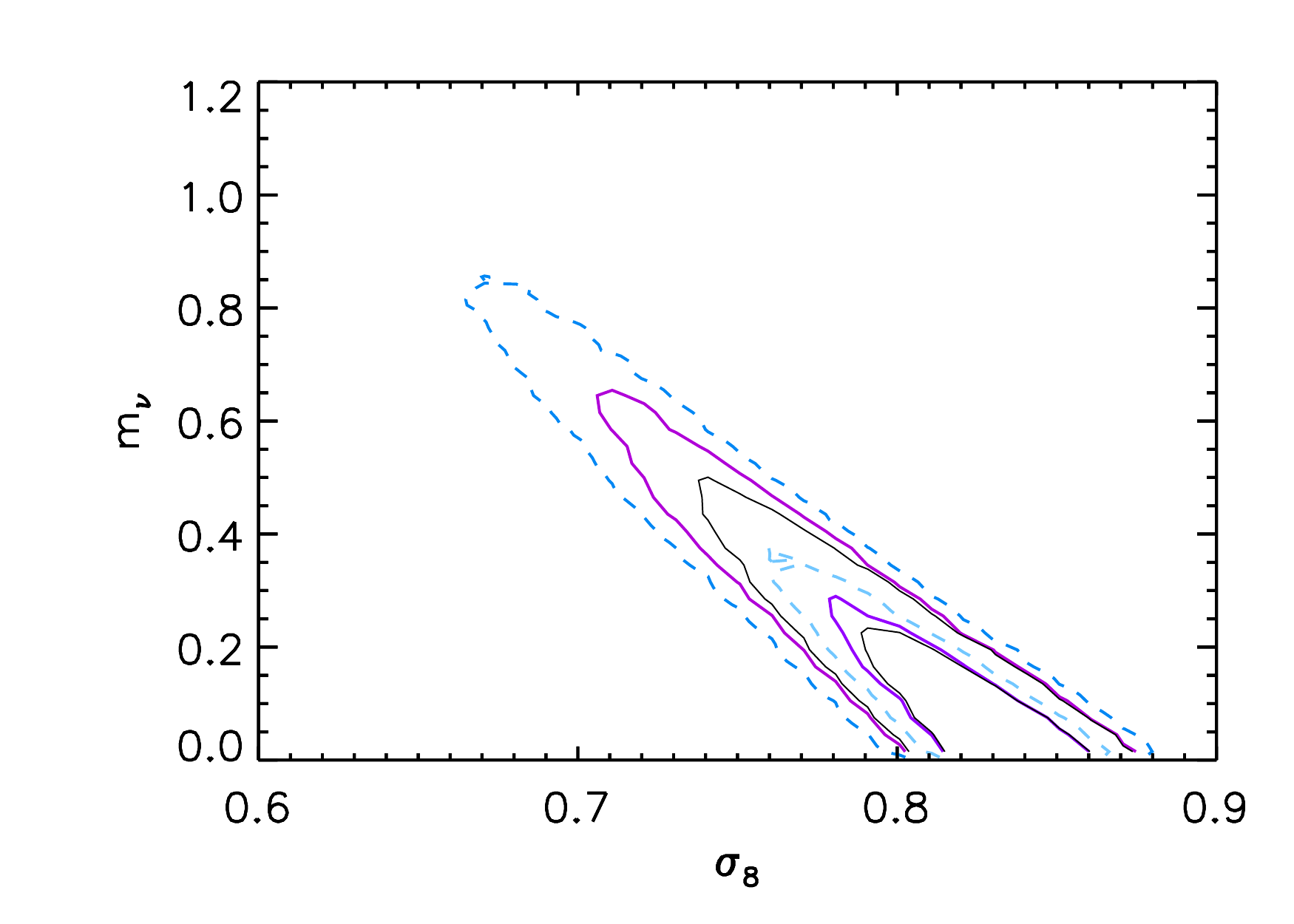}
\caption{Constraints in the $\sigma_8$--$m_{\nu}$ plane.  In both panels the   blue, dashed contours are the (joint) 68 and 95\% confidence regions  for {\it Planck}.  On the top panel    {\it Planck} temperature data and WMAP low $\ell$ polarisation are used ({\it Planck +WP}), while on the bottom panel also the {\it highL} data are included ({\it Planck +WP+highL}). The solid  contours  show how constraints improve by including \combonospace: thick, purple  lines correspond to the conservative estimate and thin, black lines to the optimistic one.}
\label{fig2}
\end{figure}

Massive neutrinos affect the growth of cosmological structure by suppressing clustering below their free streaming length; as a result in models with massive neutrinos the power spectrum amplitude at large-scale structure  scales  is lower than that at CMB scales.
If we allow the three standard-model neutrinos to have a non-zero mass and the sum of the masses $m_{\nu}$ to be the parameter to be  constrained, the {\it Planck+WP} data  constraints are $m_{\nu}<1.31\,{\rm eV}$  at 95\% confidence,   which become $m_{\nu}<0.66\,{\rm eV}$ when the {\it highL} data are considered. Including the  \combo measurement produces  $m_{\nu}<0.68 (0.47)\,{\rm eV}$ and $m_{\nu}<0.49 (0.38)\,{\rm eV}$, respectively,  always at 95\% confidence,  which represent a factor 2 (2.8) and 1.3 (1.7) improvement, respectively. 
When the information about lensing is included (through the four point function of the CMB temperature) in the CMB analysis the constraint on neutrino masses relaxes to, $m_{\nu}<0.85\,{\rm eV}$  (95\% confidence)\footnote{This point is discussed at length in the literature and in \citet{Planck_cosmology} and is possibly due to a mild tension between the CMB damping tail  and the four-point function constraints on  the magnitude of the lensing signal.}. Including \combo brings back the upper limit to  $m_{\nu}< 0.67 (0.50)\,{\rm eV}$. Fig.~\ref{fig2} presents the constraints in the $\sigma_8-m_{\nu}$ plane and illustrates the above features. Basically the \combo measurement, by  effectively constraining $\sigma_8$, breaks the $m_{\nu}-\sigma_8$ degeneracy.

\begin{figure}
\center
\includegraphics[trim= 40 180 0 200,clip,scale=0.43]{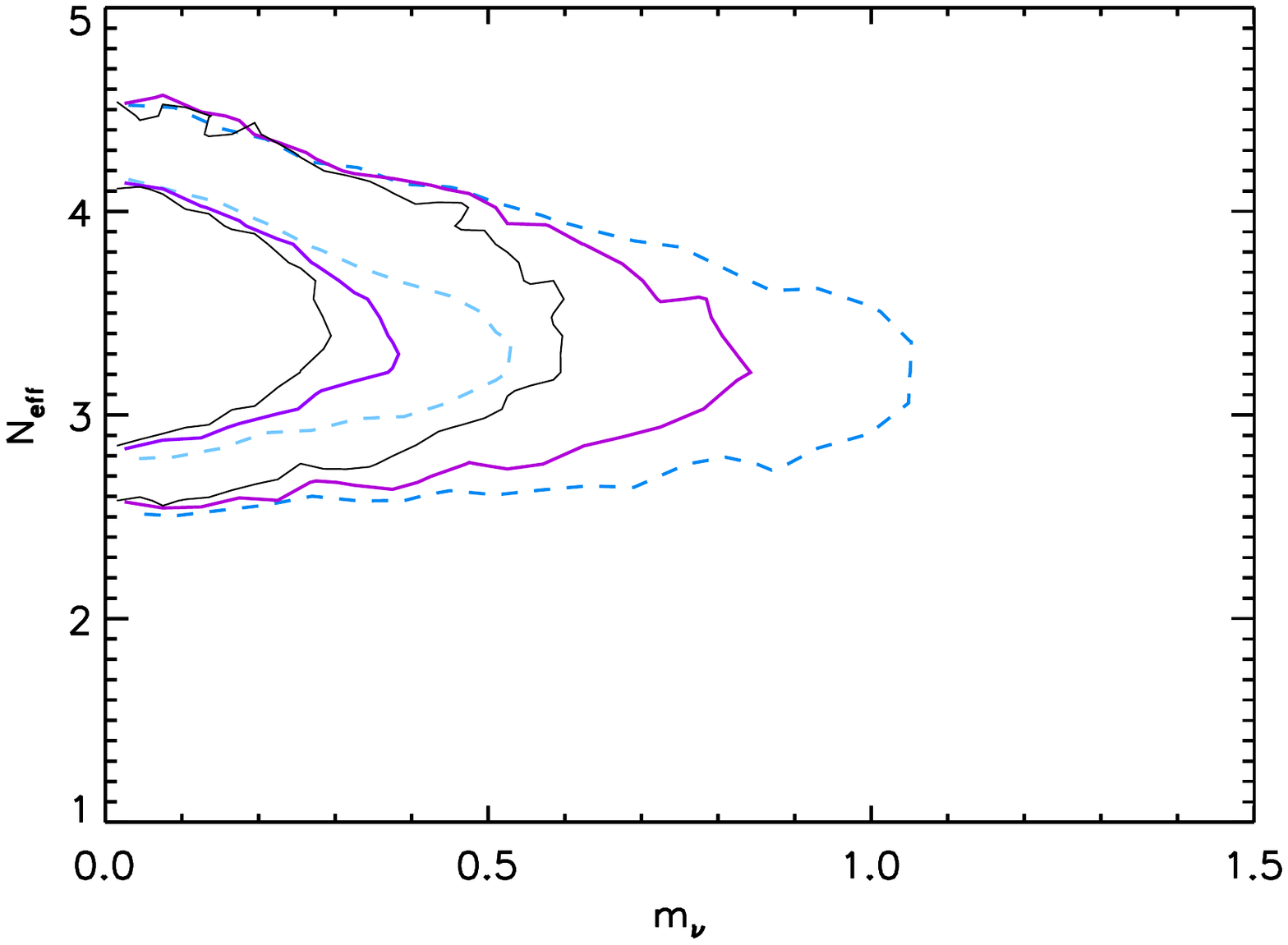}
\caption{Constraints in the total neutrino mass $m_{\nu}$, number of effective neutrino species $N_{\rm eff}$ from {\it Planck+WP} in (light) blue, dashed lines indicating 68\% and 95\% and with the addition of the \combo constraints in solid thick purple (solid thin black) contours. As above the tighter constraints are obtained with the optimistic measurement.  For $m_{\nu}$ (marginalised over $N_{\rm eff}$) we obtain that the CMB constraint $m_{\nu}<0.85\,{\rm eV}$ (at 95\% confidence) becomes $m_{\nu}<0.63 (0.46)\,{\rm eV}$ (at 95\% confidence).}
\label{fig3}
\end{figure}

A slightly more general extension of the $\Lambda$CDM model  is the case where both the number of effective neutrino species $N_{\rm eff}$ and the neutrino mass $m_{\nu}$ are treated as free parameters. Also  in this case  the \combo measurement improves the constraints, especially on the sum of neutrino masses. This behaviour is illustrated in Fig.~\ref{fig3}, where the blue, dashed contours are for {\it Planck+WP} and the solid, thick purple (thin, black) contours are in combination with  the \combo measurement. For $m_{\nu}$ (marginalised over $N_{\rm eff}$) the CMB constraint $m_{\nu}<0.85\,{\rm eV}$ (at 95\% confidence) becomes $m_{\nu}<0.63 (0.46)\,{\rm eV}$ (at 95\% confidence).

Qualitatively similar results are also obtained for the massive sterile neutrino case, where the extra sterile neutrinos are made massive rather than having the mass being equally distributed among all neutrino families:   $m_{\nu}<0.51(0.42)\,{\rm eV}$ at 95\% confidence when  we  impose a  limit to the physical thermal mass for the sterile neutrino $<10\,{\rm eV}$,  for which   the particles are distinct from cold or warm dark matter (see Table \ref{tab:1} for details). 
Some large-scale structure datasets, including the  cluster abundance from the {\it Planck } Sunyaev-Zeldovich clusters \citep{PlanckSZcosmology}, yield a much lower value for $\sigma_8 (z=0)$ than that inferred from the CMB assuming a standard $\Lambda$CDM-model with near massless neutrinos. This mismatch has been interpreted as an evidence of non-zero neutrino mass with  $m_{\nu}\sim0.45\,{\rm eV}$.  In particular 
the joint analysis of {\it Planck} temperature data with the cluster abundance from the {\it Planck } Sunyaev-Zeldovich clusters \citep{PlanckSZcosmology} yields a tentative detection of neutrino masses $m_{\nu}=0.45-0.58 \pm 0.21\,{\rm eV}$ depending on  assumptions about the calibration of the mass-observable relation. The  \combo measurement seems to disfavour the ``new physics in the neutrino sector" interpretation of the $\sigma_8$ mismatch.
\begin{table*}
\begin{center}
\begin{tabular}{|cc c c||c||c|}
\hline
& \multicolumn{3}{|c|}{$m_{\nu}-\Lambda$CDM} &$N_{\rm eff}-m_{\nu}-\Lambda$CDM&  $N_{\rm eff}-m_{\rm eff}^{\rm sterile}-\Lambda$CDM\\
\hline
&{\it Planck +WP}& {\it Planck +WP+highL}& {\it Planck +WP+highL+lensing}& {\it Planck+WP}& {\it Planck +WP+highL}\\
&$m_{\nu}<1.31\,{\rm eV}$ & $m_{\nu}<0.66\,{\rm eV}$ & $m_{\nu}<0.85\,{\rm eV}$ & $m_{\nu}<0.85\,{\rm eV}$ & $m_{\nu}<0.59\,{\rm eV}$ \\
\hline
&+\combo&+\combo&+\combo&+\combo&+\combo\\
{\rm conserv.}&$m_{\nu}<0.68\,{\rm eV}$ &$m_{\nu}<0.49\,{\rm eV}$ &$m_{\nu}<0.67\,{\rm eV}$ & $m_{\nu}<0.63,{\rm eV}$ & $m_{\nu}<0.51\,{\rm eV}$\\
{\rm optim.}&$m_{\nu}<0.46\,{\rm eV}$ &$m_{\nu}<0.38\,{\rm eV}$ &$m_{\nu}<0.50\,{\rm eV}$ & $m_{\nu}<0.46\,{\rm eV}$ & $m_{\nu}<0.42\,{\rm eV}$\\
\hline
 \end{tabular}
\end{center}
\caption{Constraints (95\% limits) on the sum of neutrino masses for several models and data set combinations. The  $m_{\nu}-\Lambda$CDM model is a  spatially flat power law $\Lambda$CDM model where the sum of neutrino masses is an extra parameter. The  $N_{\rm eff}-m_{\nu}-\Lambda$CDM model is  a  spatially flat power law $\Lambda$CDM model where both the effective number of neutrino species and the total neutrino mass are  extra parameters. The $N_{\rm eff}-m_{\rm eff}^{\rm sterile}-\Lambda$CDM model is similar to $N_{\rm eff}-m_{\nu}-\Lambda$CDM,  but where the massive neutrinos are only the sterile ones. To calculate the constraints we have imposed a physical thermal mass for the sterile neutrino $<10\,{\rm eV}$, which  defines the  region  (for the CMB) where the particles are distinct from cold or warm dark matter.  }
\label{tab:1}
\end{table*}
These results are summarised in Table \ref{tab:1}.

\begin{figure*}
\center
\includegraphics[scale=0.71]{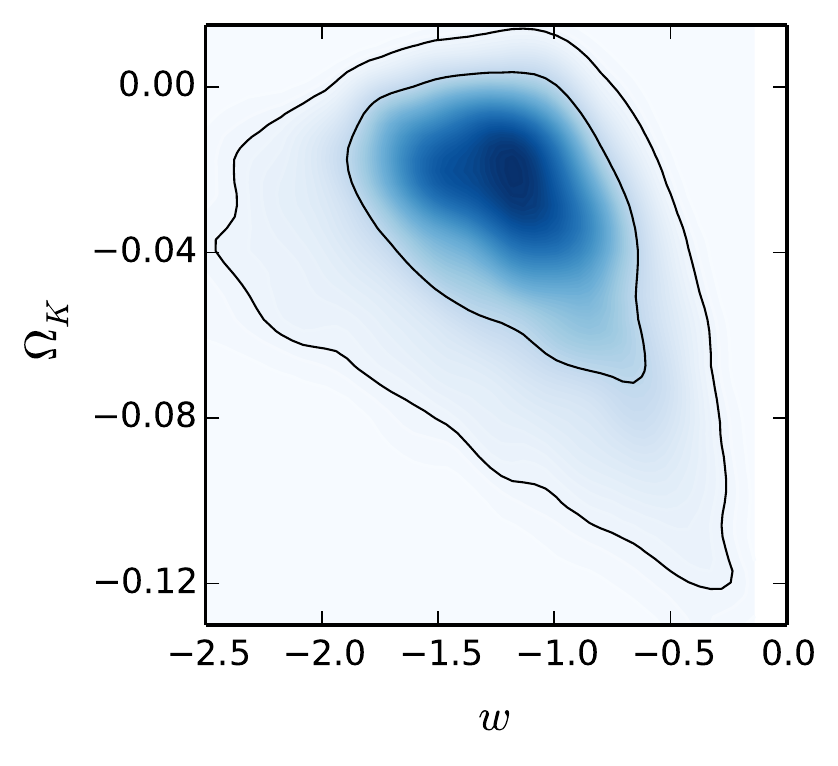}
\includegraphics[scale=0.71]{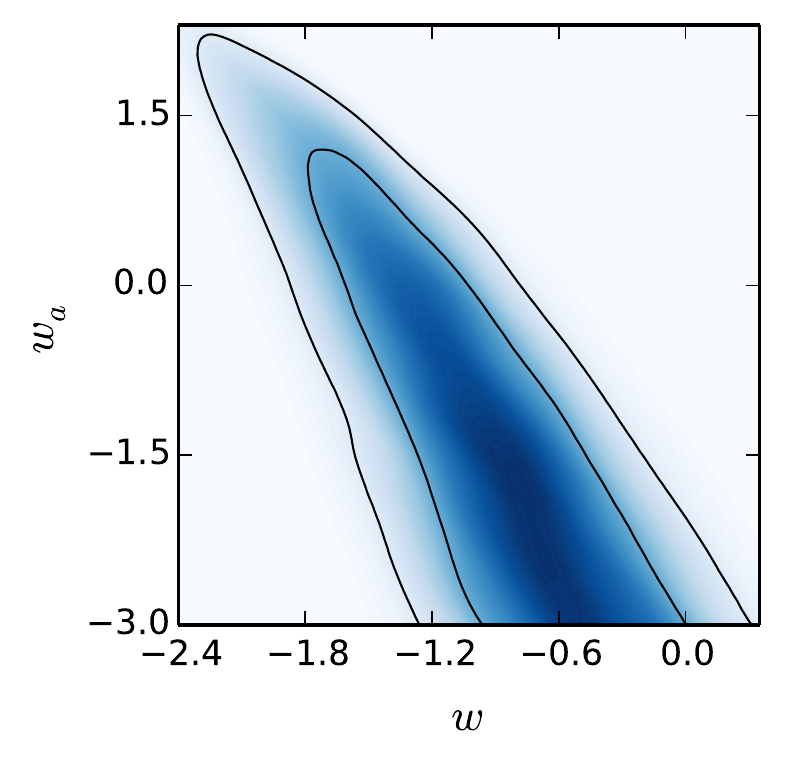}
\includegraphics[scale=0.71]{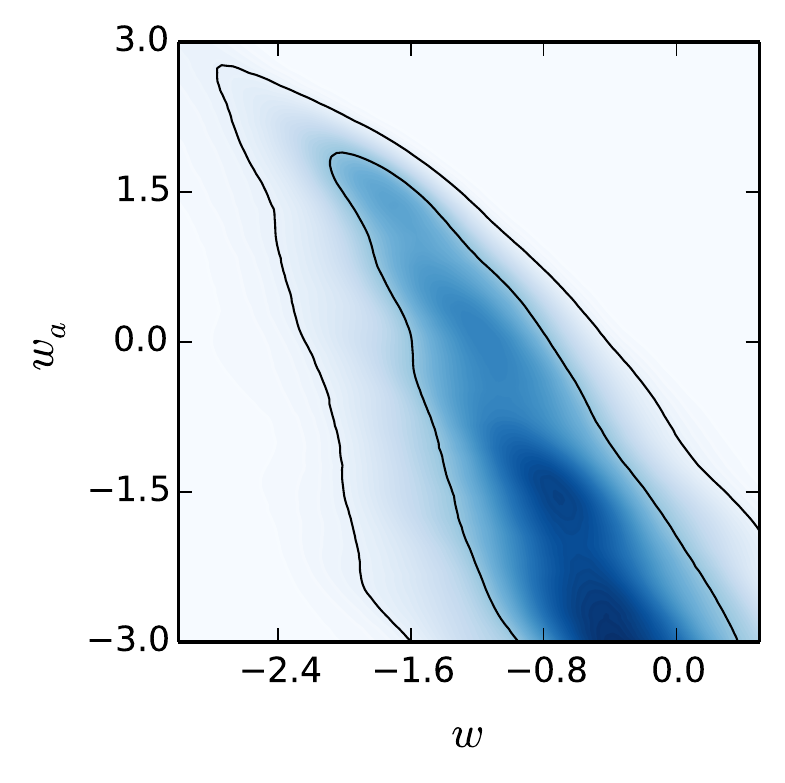}
\caption{Constraints obtained from {\it Planck+WP} in combination with the \combo measurement for the following models. In the left panel constraints in the $\Omega_k$-$w$ plane for  a non-flat Universe where the dark energy equation of state is constant but not  necessarily $-1$; in the middle panel  constraints in the $w$-$w_a$ plane for a flat model where the dark energy equation of state parameter changes with the scale factor $a$ as $w(a)=w+w_a(1-a)$. The right panel is the same as the middle panel but where the spatial flatness assumption is relaxed. In this case   $-0.076<\Omega_k<0.009$ (95\% confidence), respectively. The contour lines represent the 68\% and the 95\% confidence regions. The saturation of the colour is proportional to the posterior likelihood.}
\label{fig:w}
\end{figure*}

\subsection{Dark energy equation of state constraints}
\label{sec:3.2}
In the case of a non-flat model where the dark energy equation of state parameter $w$ is constant, but not necessarily equal to $-1$ --$ow$CDM--, the  combination of {\it Planck+WP} and \combo measurements constrain $w$ to be $-2.10<w<-0.33$ ($-1.94<w<-0.62$)  at 95\% confidence and the curvature to be $-0.093<\Omega_k<+0.008$ ($-0.076<\Omega_k<+0.007$), also at 95\% confidence. The joint constraints in the $\Omega_k$--$w$ plane are displayed in the left panel of  Fig.~\ref{fig:w}.

Conversely, if we assume flat geometry, but allow the dark energy equation of state to change with the scale-factor $a$ (according  to \citealt{wwa1} and \citealt{Linder}) as $w(a)=w+w_a(1-a)$ --$ww_a$CDM--, we obtain the constraints presented in  the middle panel of Fig.~\ref{fig:w}. The single parameter constraints are:  $-2.03<w<-0.06$ ($-1.80<w<-0.16$)  and $w_a<1.27 (1.08)$ (at 95\% confidence). These constraints do not degrade significantly if   flatness is relaxed --$oww_a$CDM--, as shown in the right panel of Fig.~\ref{fig:w}. In this case the constraint on the geometry becomes  $-0.083<\Omega_k<0.007$($-0.074<\Omega_k<0.007$) at 95\% confidence and for the dark energy parameters  $-2.38<w< 0.39$ ($-2.20<w<-0.01$) and  $w_a<1.64 (1.60)$ (95\% confidence).
For all these cases,   we ran  new Markov Chains rather than  importance sampling  existing ones. We conclude that a dark energy component is needed and is dominant  even in non-flat  models where the dark energy equation of state parameter is not necessarily constant.  The density of dark energy in units of the critical density $\Omega_{\rm dark\, energy}$ at 68\% confidence  is $0.61\pm 0.13$ ($0.637\pm0.090$) in the $ow$CDM model,  $0.742 \pm 0.071$ ($0.728\pm0.055$) in the $ww_a$CDM model  and $0.62 \pm 0.12$ ($0.639\pm 0.086$) in  the $o w w_a$CDM model.  These constraints are obtained using only data at $z\geq 0.57$ (i.e., \combo and CMB). Any more ``local" explanation for dark energy is therefore  disfavoured. 

\subsection{Modifications of GR}
\label{sec:3.3}
In modern cosmology the rationale behind introducing modifications of GR is to explain the late-time cosmological acceleration. Therefore,  
the most popular modifications of GR  mimic the effects of the cosmological constant on the expansion history and  become important only at low redshifts.  If we allow gravity to deviate from  GR,  the CMB  offers  only weak  constrains on the late-time growth of structures, via the Integrated Sachs-Wolfe (ISW) effect and lensing.
A popular parametrisation for deviations from GR growth is given by,
\begin{equation}
f(z)=\Omega_m(z)^\gamma.
\label{eq:gamma}
\end{equation}
For $\Lambda$CDM+GR $\gamma|_{\Lambda{\rm CDM}}\simeq 0.56$; for  dark energy models with an equation of state parameter different from $w=-1$, $\gamma$ acquires a  weak redshift dependence and its value does not deviate significantly from $\gamma|_{\Lambda{\rm CDM}}$. 
Since we have a  measurement at a given redshift  we consider $\gamma$ to be constant.  We  also assume that this modification affects the late-time Universe and not the CMB.  This assumption is reasonable as  in this parametrisation, as $\Omega_m(z)\longrightarrow 1$ (i.e., for most cosmologies at $z\gg 1$) the growth becomes that of an Einstein-De-Sitter Universe for any value of $\gamma$.

For this extension to the base model we assume that  the background expansion history is given by that of the $\Lambda$CDM-model as constrained by {\it Planck+WP} (using  {\it Planck+WP+ highL} +BAO does not change the results significantly). The constraints on the growth rate reduce to $\gamma=0.40^{+0.50}_{-0.42}$ ($^{+0.28}_{-0.26}$) at 68\% confidence and $\gamma <0.87$ ($0.80$) at 95\% confidence.

For coupled dark energy-dark matter models the growth  can be parametrised as (e.g., \cite{diportoamendola} and references therein),
\begin{equation}
f(z)=\Omega_m(z)^{0.56}(1+\eta).
\label{eta_para}
\end{equation}
 Using this equation we obtain
$\eta=0.055\pm 0.145$ $(\pm 0.090 $) at 68\% confidence. For the coupled dark energy-dark matter models, $\eta$ is related to the coupling constant $\beta_c$. These models  have a non-negligible amount of dark energy at early times, so they can be constrained by the CMB. Nevertheless,   the $\eta$ constraint can be re-interpreted  as a limit on the coupling constant
 $\beta_c<0.34$ ($<0.28$) at 95\% confidence at the effective survey redshift $z=0.57$; as expected   this constraint is much weaker than that obtained from the CMB by \cite{Pettorinoetal:2012} assuming a constant coupling.

Inspired by \cite{Acquavivaetal08}, who introduced the quantity $\epsilon=f/f|_{\Lambda{\rm CDM}}-1$, we can define
\begin{equation}
\epsilon^{\prime}\equiv\frac{f^{0.43}\sigma_8}{f^{0.43}\sigma_8|_{\Lambda{\rm CDM}}}-1,
 \label{eq:epsilon}
\end{equation}
which is a more model-independent approach than the  $\gamma$-parameterisation of Eq. \ref{eq:gamma} or the  $\eta$-parameterisation of Eq. \ref{eta_para}. Eq. \ref{eq:epsilon} also  enables one to quantify possibly early times  deviations from GR\footnote{In the $\gamma$ parameterisation the standard growth is recovered for all values of $\gamma$ at $z> 0.5$, when $\Omega_m\rightarrow 1$.}. Note that since $\sigma_8(z) / \sigma_8(z)|_{\Lambda{\rm CDM}}=D(z)/D(z)|_{\Lambda{\rm CDM}} $, where $D(z)$ is the linear growth factor, Eq. \ref{eq:epsilon} parametrizes deviations on the \combo (or on the $f^{0.43}D|_{0.57}$) produced by changes in the theory of gravity on $f$ {\it and} $D$, whereas Eq. \ref{eta_para} only accounts for changes on $f$.
This quantity, which is identical to zero for $\Lambda$CDM and exceedingly close to zero for minimally coupled quintessence-type models, is redshift dependent and can, in principle, be also scale dependent when departures from GR are present. Here we  assume $\epsilon^{\prime}$ is scale-independent over the scales probed and we compute its value at the  effective survey redshift.
 For $f^{0.43}\sigma_8|_{\Lambda{\rm CDM}}$ we take the range predicted  by the {\it Planck+WP}  combination and obtain, at 68\% confidence,
 \begin{equation}
 \epsilon^{\prime}(z=0.57)=0.04\pm 0.15 (\pm 0.10)\,,
 \end{equation}
 in  agreement with the GR value of zero.
 
 \section{Breaking the \lowercase{$f$}-$\sigma_8$ degeneracy} \label{sec:PoP}
As  displayed in Fig. \ref{fig1}, the constraints in the $f-\sigma_8$ plane produced by the  redshift-space distortions of the anisotropic redshift-space  correlation function and those produced by the monopole of the power spectrum and bispectrum are highly complementary.
 
 A joint analysis combining the two-point anisotropic clustering and the bispectrum monopole is able to break the degeneracy between $f$ and $\sigma_8$, enabling the measurement of both quantities separately. Here we do not attempt a full, combined analysis of the power spectrum monopole, quadrupole and the bispectrum monopole, which would yield the combined constraint of multiples parameters, including bias if a single consistent bias model were adopted. We will consider such analysis in a future work.  
 
\begin{figure}
 \center
 \includegraphics[scale=0.33]{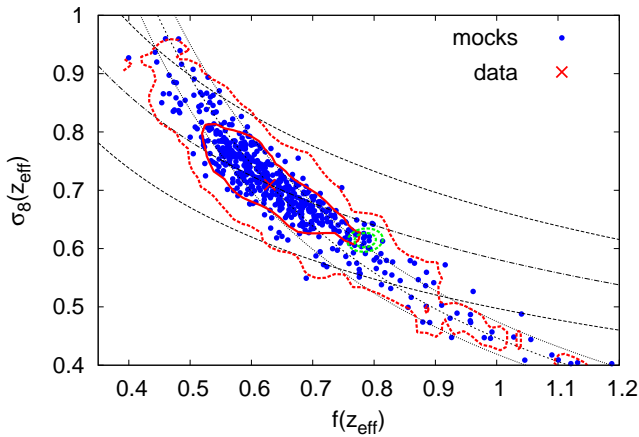}
\caption{Constraints in the $f$--$\sigma_8$ plane where both quantities are at the effective redshift of the BOSS CMASS DR11 galaxies $z=0.57$. In this figure, for \combo we consider the conservative measurement only. The blue dots represent the best-fitting values for 600 mocks when the results from \citet{Samushiaetal:14} (short dashed line) and Paper~I (dot-dashed line) are combined (see text for details). The black dotted and dashed lines show the $1\sigma$ errors, respectively. The contours correspond to the  68 and 95\% confidence regions (solid and dashed lines, respectively) estimated from the dispersion of the mocks.  The green dashed ellipses represent the Planck CMB inferred constraints (68 and 95\% confidence) when assuming GR and a $\Lambda$CDM model. The red cross represents the best-fitting value for the BOSS CMASS DR11 data. The best-fitting values of the galaxy mocks, as well as the contours, have been displaced in the logarithmic space to be centered on  the  measurement from the data.}

\label{fig:PoP}
 \end{figure}

Instead, we perform a combined, {\it a posteriori}, analysis between the measurements of $f\sigma_8|_{z=0.57}$ obtained by \cite{Samushiaetal:14},  when the background expansion
 is fixed to the one predicted by Planck,  and of $f^{0.43}\sigma_8|_{z=0.57}$ obtained  in Paper~I. Although the  measurements were carried out  independently, and one is performed in configuration  space while the other in Fourier space, they share the information related to the  monopole of the two-point statistics, and therefore they are expected to be moderately correlated. Using measurements based on the same set of mocks \citep{manera} we  compute the  correlation and errors from their dispersion (see \S~3.9 in Paper~I  for details of the method).  We consider the measurement of $[f\sigma_8]_i$ and $[f^{0.43}\sigma_8]_i$ for each single $i$~-~mock, and combining them we  extract the corresponding values:
 \begin{eqnarray}
 [f]_i&=& \left\{[f\sigma_8]_i/[f^{0.43}\sigma_8]_i\right\}^{1/0.57},\\
 {[\sigma_8]}_i&=& [f\sigma_8]_i/[f]_i.
 \end{eqnarray}
 The errors on $f$ and $\sigma_8$ are estimated from the dispersion of $[f]_i$ and $[\sigma_8]_i$ among all the mocks, respectively. This result is illustrated in Fig.~\ref{fig:PoP}, where the blue dots represent the obtained values of $[f]_i$ and $[\sigma_8]_i$ for the 600 realisations of the galaxy mocks: $i=1,2,\ldots, 600$.
   
 The combined constraints in the $f-\sigma_8$ plane are displayed in Fig.~\ref{fig:PoP}, which illustrates the constraints on $f\sigma_8$ and $f^{0.43}\sigma_8$ from \cite{Samushiaetal:14} and Paper~I,  respectively, using the same line-notation as in Fig.~\ref{fig1}.  Blue dots represent the best-fit values for the mocks and the red cross shows the best-fit value for the DR11 CMASS dataset. The green dashed ellipses represent the Planck CMB inferred constraints (68 and 95\% confidence) when assuming GR and a $\Lambda$CDM model. The red contours correspond to the  (joint)  68\% (solid lines) and 95\% (dashed lines) confidence levels extracted from the mocks and centered on the data,  where the mild prior $0<f<2$ has been used.
 
Originally the constrains on $f\sigma_8$ and $f^{0.43}\sigma_8$ were: $f\sigma_8|_{z=0.57}=0.447\pm0.028$ and $f^{0.43}\sigma_8|_{z=0.57}=0.582\pm0.084\,(0.051)$. After combining the measurements  the degeneracy between $f$ and $\sigma_8$ is broken, although the two parameters remain significantly (anti)correlated, with a correlation coefficient of $\sim-0.9$. From the joint distribution we can now  obtain marginalised constraints on each of the parameters: $f(z_{\rm eff})=  0.63 \pm 0.16 (0.13)$ (marginalised over $\sigma_8$), $\sigma_8(z_{\rm eff})=0.710\pm 0.086(0.069)$ (marginalised over $f$), where all the reported errors are at the 68\% confidence level. These values represent a $12 (10)\%$ and $25 (20)\%$ relative error for $\sigma_8(z_{\rm eff})$ and $f(z_{\rm eff})$, respectively. 

 Previous works in the literature  have used  bispectrum alone or in combination with the  power spectrum  of galaxies to break degeneracies present in the power spectrum analysis.  \citet{scoccimarro_IRAS,Verdeetal:2002,Nishimichietal:2007,McBrideetal:2011,chiang15} constrain  the bias parameters $b_1$ and $b_2$ with the bisectrum (or equivalent observables), the derived $b_1$ parameter can then be used in combination with the power spectrum-derived $\beta \sim f/b_1$ to constrain $f$.   \cite{Marinetal:2013} use the bispectrum to constrain the amplitude of primodrial fluctuations $\sigma_8$ when other cosmological parameters (such as $f$) were fixed to fiducial  ($\Lambda$CDM, GR) values. However, this is the first time that these two quantities $f$ and $\sigma_8$ have been separately determined from galaxy clustering using  the power spectrum and bispectrum measurements.

The  resulting values for $f$ and $\sigma_8$ are in broad agreement with the CMB-inferred values. This can be appreciated in Fig. \ref{fig:PoP}, where the green dashed ellipses are in general agreement with the red contours of the data.  We have also computed the tension ${\cal T}$ as introduced in \cite{Local}, which quantifies whether multi-dimensional cosmological parameters constraints from two different experiments are in agreement or  not. We find that the tension is not significant, $\ln {\cal T}<1$, i.e., the two measurements are in agreement.

 We have repeated the analysis of \S~\ref{sec:3.1} and \ref{sec:3.2} using these new constraints, but we find that the constraints on the cosmological parameters do not change in any significant way. At the current size of the error-bars of \combonospace, the $f\sigma_8$ degeneracy is being cut for high values of $\sigma_8$, but has a tail for low values of $\sigma_8$, as it is shown in Fig.~\ref{fig:PoP}. Therefore, breaking the $f\sigma_8$ degeneracy in this way (in combination with Planck data)  does not improve significantly the error-bars in the studied parameters.  We expect that this will improve with the forthcoming analysis of DR12, where a joint analysis of power spectrum monopole, quadrupole and bispectrum monopole will be performed. 

When  the measurements obtained on $f$ are applied to the parameters of Eq. \ref{eq:gamma}  the constraints on $\gamma$ become $\gamma=0.40\pm0.43 (0.35)$ (68\% confidence). We can now consider the variable $\epsilon$ introduced by \cite{Acquavivaetal08}, obtaining ${\epsilon}=-0.21\pm0.56 (0.45)$, also at 68\% confidence. Both measurements should be considered at    $z_{\rm eff}=0.57$ at scales of $k\simeq0.1\,h{\rm Mpc}^{-1}$. In order to  be able to use these measurements to distinguish between GR and popular  and viable modifications of gravity   that  match the $\Lambda$CDM model expansion history, error-bars would have to be reduced by a factor of few.

 In this paper we only have considered to combine the results obtained from the power spectrum and bispectrum monopole analysis, with those from the anisotropic two-point correlation function \cite{Samushiaetal:14} . Moreover we only consider constraints on the parameters describing the growth of perturbations and not the expansion history. The expansion history can be probed with the  power spectrum  (both monople and quadrupole) via the parameters $D_V/r_s$ and  $F$ as it has been done by the  same collaboration \citep{Samushiaetal:14}. But exploring  the effect of adding  the bispectrum information to e.g. break the degeneracy between  $F$ and $f\sigma_8$ goes beyond the scope of this paper.
 
 It may seem disappointing at first sight that the time consuming and challenging analysis of the galaxy bispectrum does not seem to improve dramatically the cosmological constraints on parameters such as $\gamma$. However,  is important to keep in mind that   the bispectrum is a different statistic from the  monopole and quadrupole of the power spectrum, which relies on different modelling, different physical effects and  is affected by different systematics.  It adds some additional information to the power spectrum analysis which goes beyond the size of the error-bars on some cosmological parameters. It offers a consistency check. 
It gives insights on the behaviour and amplitude of the galaxy bias, it serves as a  test of our modelling of mild non-linearities and non-linear redshift space distortions. 

We have shown for example that adding our bispectrum result on $f^{0.43}\sigma_8$ to Planck  data, constrains neutrino mass as much as adding the ``{\it highL}" and ``{\it highL}" combined with  Planck {\it lensing}. 
 
 Moreover, it has been proposed that the bispectrum in combination with the power spectrum could be used to constrain not just gravity and bias, but also the nature of the initial conditions (primordial non-Gaussianity)  from future surveys. It is important to start exploring the challenges and opportunities that a combined power spectrum and bispectrum analysis offer.

\section{conclusions}
\label{sec:5}
We have examined the cosmological implications of the constraints on the quantity $f^{0.43}\sigma_8$ at $z_{\rm eff}=0.57$, which offers a direct cosmological and model-independent probe of the growth of structure. This constraint has been obtained from the measurement of the power spectrum and bispectrum monopoles of the SDSSIII BOSS DR11 CMASS galaxies and  has been recently presented in a companion paper \citep{paper1} (Paper~I). 

We have combined this result with recent state-of-the art CMB  constraints for several underlying cosmological models.
We find  agreement in the  standard $\Lambda$CDM cosmological paradigm between the growth of structure predicted by CMB observations and the direct measurement from galaxy clustering. 

When considering popular  $\Lambda$CDM model extensions, which still rely on GR,  we find that 
this new measurement is useful to improve the CMB constraints on cosmological parameters only for  model extensions   that involve massive neutrinos or  for   models where dark energy deviates from a cosmological constant and where more than one parameter is added. The \combo measurement improves CMB neutrino mass constraints  by  at least 30\% and in some cases by as much as factor 2 to 2.8. (see Table \ref{tab:1} for details). There is no evidence for non-standard neutrino properties when considering CMB and \combo measurements.

For dark energy models where the  equation of state parameter of dark energy is not constant, or for models where it is constant but not equal to $-1$ and the geometry of the Universe is not constrained to be flat,  we can obtain interesting constraints.  Curvature is constrained at the 8\% level (95\%   confidence). We find no evidence for any deviations from  a cosmological constant, but  dark energy is needed  as a dominant component of the Universe, even for non-flat, non-$\Lambda$CDM cosmologies. This conclusion is reached using  only data at $z\geq 0.57$, thus disfavouring  ``local" explanations of dark energy.

We  have also  examined the constraints that the measurement of \combonospace, in combination with data on the Universe's geometry  and expansion history, provide on modifications of GR. We  have examined different phenomenological parametrisations of how the growth of structure  is modified when we relax the assumption of GR.  We do not observe any significant tension between these measurements and GR predictions, in particular we find that $\gamma=0.40^{+0.50}_{-0.45} (^{+0.28}_{-0.26})$ (68\% confidence), where $f(z)=\Omega_m(z)^\gamma$.

Finally,  we have presented how the measurement of \combo can be combined with the measurement of $f\sigma_8|_{z=0.57}$ from the same galaxy sample to break the degeneracy between $f$ and $\sigma_8$. This is the first time that a separate measurement of $f$ and $\sigma_8$ has been obtained using  power spectrum and bispectrum measurements from galaxy clustering: $f=0.63\pm0.16$ and $\sigma_8=0.710\pm0.086$, both at $z=0.57$. The size of errors already provides an insight on how powerful a fully and optimal joint analysis can be.   We find that $f$ can be measured with a relative precision of  $\sim25\%$, and $\sigma_8$ with $\sim$10\% at $68\%$ confidence level. We expect that the size of these error-bars can be reduced if the power spectrum multipoles and bispectrum monopole are combined {\it prior} to obtain the $f$ and $\sigma_8$ best-fitting values.   Further testing for potential systematics would also be of benefit for such novel analysis.

While the $f-\sigma_8$ degeneracy could also be broken using measurements at several different redshifts, or resorting to  weak lensing data or cross correlation with the weak lensing signal of  the CMB, 
the approach described in this paper provides a complementary and self-contained approach  to achieve the same goal relying on  galaxy redshift surveys alone, without the need of wide  redshift coverage. 

\section*{Acknowledgements}
HGM is grateful for support from the UK Science and Technology Facilities Council through the grant
ST/I001204/1. 
LV   is supported by the European Research Council under the European Community's Seventh Framework Programme grant FP7-IDEAS-Phys.LSS and  acknowledges Mineco grant FPA2011-29678- C02-02. 
JN is supported in part by ERC grant FP7-IDEAS-Phys.LSS.
AC is  supported by the European Research Council under the European Community's Seventh Framework Programme grant FP7-IDEAS-Phys.LSS.
WJP is grateful for support from the UK Science and Technology Facilities Research Council through the grant 
ST/I001204/1, and the European  Research Council through the grant ``Darksurvey".

Funding for SDSS-III has been provided by the Alfred P. Sloan
Foundation, the Participating Institutions, the National Science
Foundation, and the U.S. Department of Energy Office of Science. The
SDSS-III web site is http://www.sdss3.org/.

SDSS-III is managed by the Astrophysical Research Consortium for the
Participating Institutions of the SDSS-III Collaboration including the
University of Arizona,
the Brazilian Participation Group,
Brookhaven National Laboratory,
University of Cambridge,
Carnegie Mellon University,
University of Florida,
the French Participation Group,
the German Participation Group,
Harvard University,
the Instituto de Astrofisica de Canarias,
the Michigan State/Notre Dame/JINA Participation Group,
Johns Hopkins University,
Lawrence Berkeley National Laboratory,
Max Planck Institute for Astrophysics,
Max Planck Institute for Extraterrestrial Physics,
New Mexico State University,
New York University,
Ohio State University,
Pennsylvania State University,
University of Portsmouth,
Princeton University,
the Spanish Participation Group,
University of Tokyo,
University of Utah,
Vanderbilt University,
University of Virginia,
University of Washington,
and Yale University.
This research used resources of the National Energy Research Scientific
Computing Center, which is supported by the Office of Science of the
U.S. Department of Energy under Contract No. DE-AC02-05CH11231.

Results are based on observations obtained with Planck (http://www.esa.int/Planck), an ESA science mission with instruments and contributions directly funded by ESA Member States, NASA, and Canada.

Numerical computations were done on Hipatia ICC-UB BULLx High Performance Computing Cluster at the University of Barcelona.

%
%
%


\def\jnl@style{\it}
\def\aaref@jnl#1{{\jnl@style#1}}

\def\aaref@jnl#1{{\jnl@style#1}}

\def\aj{\aaref@jnl{AJ}}                   
\def\araa{\aaref@jnl{ARA\&A}}             
\def\apj{\aaref@jnl{ApJ}}                 
\def\apjl{\aaref@jnl{ApJ}}                
\def\apjs{\aaref@jnl{ApJS}}               
\def\ao{\aaref@jnl{Appl.~Opt.}}           
\def\apss{\aaref@jnl{Ap\&SS}}             
\def\aap{\aaref@jnl{A\&A}}                
\def\aapr{\aaref@jnl{A\&A~Rev.}}          
\def\aaps{\aaref@jnl{A\&AS}}              
\def\azh{\aaref@jnl{AZh}}                 
\def\baas{\aaref@jnl{BAAS}}               
\def\jrasc{\aaref@jnl{JRASC}}             
\def\memras{\aaref@jnl{MmRAS}}            
\def\mnras{\aaref@jnl{MNRAS}}             
\def\pra{\aaref@jnl{Phys.~Rev.~A}}        
\def\prb{\aaref@jnl{Phys.~Rev.~B}}        
\def\prc{\aaref@jnl{Phys.~Rev.~C}}        
\def\prd{\aaref@jnl{Phys.~Rev.~D}}        
\def\pre{\aaref@jnl{Phys.~Rev.~E}}        
\def\prl{\aaref@jnl{Phys.~Rev.~Lett.}}    
\def\pasp{\aaref@jnl{PASP}}               
\def\pasj{\aaref@jnl{PASJ}}               
\def\qjras{\aaref@jnl{QJRAS}}             
\def\skytel{\aaref@jnl{S\&T}}             
\def\solphys{\aaref@jnl{Sol.~Phys.}}      
\def\sovast{\aaref@jnl{Soviet~Ast.}}      
\def\ssr{\aaref@jnl{Space~Sci.~Rev.}}     
\def\zap{\aaref@jnl{ZAp}}                 
\def\nat{\aaref@jnl{Nature}}              
\def\iaucirc{\aaref@jnl{IAU~Circ.}}       
\def\aplett{\aaref@jnl{Astrophys.~Lett.}} 
\def\apspr{\aaref@jnl{Astrophys.~Space~Phys.~Res.}}
\def\bain{\aaref@jnl{Bull.~Astron.~Inst.~Netherlands}} 
\def\fcp{\aaref@jnl{Fund.~Cosmic~Phys.}}  
\def\gca{\aaref@jnl{Geochim.~Cosmochim.~Acta}}   
\def\grl{\aaref@jnl{Geophys.~Res.~Lett.}} 
\def\jcp{\aaref@jnl{J.~Chem.~Phys.}}      
\def\jgr{\aaref@jnl{J.~Geophys.~Res.}}    
\def\jqsrt{\aaref@jnl{J.~Quant.~Spec.~Radiat.~Transf.}}
\def\memsai{\aaref@jnl{Mem.~Soc.~Astron.~Italiana}}
\def\nphysa{\aaref@jnl{Nucl.~Phys.~A}}   
\def\physrep{\aaref@jnl{Phys.~Rep.}}   
\def\physscr{\aaref@jnl{Phys.~Scr}}   
\def\planss{\aaref@jnl{Planet.~Space~Sci.}}   
\def\procspie{\aaref@jnl{Proc.~SPIE}}   
\def\jcap{\aaref@jnl{J. Cosmology Astropart. Phys.}}

\let\astap=\aap
\let\apjlett=\apjl
\let\apjsupp=\apjs
\let\applopt=\ao

\newcommand{\etal}{et al.\ }

\newcommand{\mpc}{\, {\rm Mpc}}
\newcommand{\kpc}{\, {\rm kpc}}
\newcommand{\hmpc}{\, h^{-1} \mpc}
\newcommand{\ihmpc}{\, h\, {\rm Mpc}^{-1}}
\newcommand{\ikms}{\, {\rm s\, km}^{-1}}
\newcommand{\kms}{\, {\rm km\, s}^{-1}}
\newcommand{\hkpc}{\, h^{-1} \kpc}
\newcommand{\lya}{Ly$\alpha$\ }
\newcommand{\lyb}{Lyman-$\beta$\ }
\newcommand{\lyaf}{Ly$\alpha$ forest}
\newcommand{\lr}{\lambda_{{\rm rest}}}
\newcommand{\bF}{\bar{F}}
\newcommand{\bS}{\bar{S}}
\newcommand{\bC}{\bar{C}}
\newcommand{\bB}{\bar{B}}
\newcommand{\vdF}{{\mathbf \delta_F}}
\newcommand{\vdS}{{\mathbf \delta_S}}
\newcommand{\vdf}{{\mathbf \delta_f}}
\newcommand{\vdn}{{\mathbf \delta_n}}
\newcommand{\vdC}{{\mathbf \delta_C}}
\newcommand{\vdX}{{\mathbf \delta_X}}
\newcommand{\xrei}{x_{rei}}
\newcommand{\lrmin}{\lambda_{{\rm rest, min}}}
\newcommand{\lrmax}{\lambda_{{\rm rest, max}}}
\newcommand{\lmin}{\lambda_{{\rm min}}}
\newcommand{\lmax}{\lambda_{{\rm max}}}
\newcommand{\hi}{\mbox{H\,{\scriptsize I}\ }}
\newcommand{\heii}{\mbox{He\,{\scriptsize II}\ }}
\newcommand{\vp}{\mathbf{p}}
\newcommand{\vq}{\mathbf{q}}
\newcommand{\vxperp}{\mathbf{x_\perp}}
\newcommand{\vkperp}{\mathbf{k_\perp}}
\newcommand{\vrperp}{\mathbf{r_\perp}}
\newcommand{\vx}{\mathbf{x}}
\newcommand{\vy}{\mathbf{y}}
\newcommand{\vk}{\mathbf{k}}
\newcommand{\vR}{\mathbf{r}}
\newcommand{\tdtwo}{\tilde{b}_{\delta^2}}
\newcommand{\tstwo}{\tilde{b}_{s^2}}
\newcommand{\tbthree}{\tilde{b}_3}
\newcommand{\tadtwo}{\tilde{a}_{\delta^2}}
\newcommand{\tastwo}{\tilde{a}_{s^2}}
\newcommand{\tabthree}{\tilde{a}_3}
\newcommand{\vnabla}{\mathbf{\nabla}}
\newcommand{\tpsi}{\tilde{\psi}}
\newcommand{\vv}{\mathbf{v}}
\newcommand{\fnl}{{f_{\rm NL}}}
\newcommand{\tfnl}{{\tilde{f}_{\rm NL}}}
\newcommand{\gnl}{g_{\rm NL}}
\newcommand{\orderfour}{\mathcal{O}\left(\delta_1^4\right)}
\newcommand{\SDSSPF}{\cite{2006ApJS..163...80M}}
\newcommand{\PF}{$P_F^{\rm 1D}(k_\parallel,z)$}
\newcommand\ion[2]{#1$\;${\small \uppercase\expandafter{\romannumeral #2}}}%
\newcommand\ionalt[2]{#1$\;${\scriptsize \uppercase\expandafter{\romannumeral #2}}}%
\newcommand{\vxone}{\mathbf{x_1}}
\newcommand{\vxtwo}{\mathbf{x_2}}
\newcommand{\vRot}{\mathbf{r_{12}}}
\newcommand{\cm}{\, {\rm cm}}

\bibliographystyle{mn2e}
\bibliography{paper2.bib}
                                   
\end{document}